\documentclass[aps,pra,showpacs,twocolumn,superscriptaddress,a4paper]{revtex4}
\usepackage{bm}
\usepackage{epsfig}
\usepackage{times}
\usepackage{amssymb}
\usepackage{graphics}
\usepackage{amsmath}
\usepackage{color}

\newcommand{\bra}[1]{\ensuremath{\left<#1\right|}}
\newcommand{\ket}[1]{\ensuremath{\left|#1\right>}}

\begin{document}

\title{Coherent adiabatic theory of two-electron quantum dot molecules in external spin baths}

\author{R. Nepstad}
 \affiliation{Department of Physics and Technology, University
 of Bergen, N-5007 Bergen, Norway}

\author{L. S{\ae}len}
\affiliation{Department of Physics and Technology, University
 of Bergen, N-5007 Bergen, Norway}

\affiliation{Laboratoire de Chimie Physique-Mati\' ere et
Rayonnement, Universit\' e Pierre et Marie Curie, 11, Rue Pierre et
Marie Curie 75231 Paris Cedex 05, France}

\author{J. P. Hansen}
\affiliation{Department of Physics and Technology, University
 of Bergen, N-5007 Bergen, Norway}

\begin{abstract}
We derive an accurate molecular orbital based expression for the coherent
time evolution of a two-electron wave function in a quantum dot molecule
where the electrons interact with each other, with external time dependent
electromagnetic fields and with a surrounding nuclear spin reservoir.
The theory allows for direct numerical modeling of the decoherence in quantum
dots due to hyperfine interactions. Calculations result in good agreement with recent singlet-triplet
 dephasing experiments by Laird \textit{et. al.}[Phys. Rev. Lett. {\bf 97}, 056801 (2006)],
 as well as analytical model calculations.
 Furthermore, it is shown that using a much faster electric switch than applied in
  these experiments will transfer the initial state to excited states where the hyperfine
  singlet-triplet mixing is negligible.
\end{abstract}


\maketitle

%
%
It is now well recognized that the hyperfine interaction
is one of the main sources of decoherence in few-electron quantum
dots. This interaction, originally considered in metals by
Overhauser more than 50 years ago~\cite{overhauser}, couples the
electronic spin states through weak nuclear spin interactions with
an order of $ \sim 10^6$ surrounding nuclei~\cite{khaetskii}.
Recently, this coupling has received considerable interest through
the demonstration of controlled single electron
manipulation~\cite{koppens1, petta1} which opens for real quantum
information processing based on electronic spin states in quantum
dots~\cite{divincenzo}. Procedures to minimize or control the
hyperfine interaction is therefore vital for the functioning of any
quantum dot based information processing technology. Such mechanisms
also contain novel aspects of spin de- and re-phasing of quantum
systems interacting with a large spin bath.

Recently it was demonstrated in experiments~\cite{koppens2, petta2}
with two-electron quantum dot molecules that the magnitude of the
hyperfine interaction is consistent with a random magnetic field
strength of a few~mT. The nuclear field-induced singlet-triplet
coupling leads to a spin dephasing of an initially prepared singlet
state within $1-10$~ns. These experiments utilize fast adiabatic
electric switching techniques which transform the ground state from
a two-electron ionic state in one dot to a covalent state with one
electron in each dot.

The experiments have been analyzed in detail theoretically based on
various model Hamiltonians~\cite{taylor,loss}: For small tunneling
coupling the effective two-electron Hilbert space amounts to the
four possible covalent spin states (a singlet and three triplet
states). This may be further reduced to two states by exposing the
molecule to a magnetic field which decouple the two states with
nonzero magnetic quantum numbers. Within this approximation it was
shown that the hyperfine interaction induce a spin saturation which
saturates sensitively as a function of the exchange coupling and the
hyperfine coupling~\cite{loss}. These predictions were confirmed
experimentally by Laird \textit{et. al.}~\cite{laird}.

In the present Letter we develop a full coherent model of the
two-electron spin dynamics which includes the hyperfine interaction
on equal footing with the time dependent external fields. The
results may be directly compared with the experimental results. The
theory not only validates the effective two-level models in the
presence of an external magnetic field, but also demonstrates
predictive saturation values in the absence of an external magnetic
field. Furthermore, it will be shown that decreasing the switching
time an order of magnitude may lead to controlled diabatic
transfer~\cite{murgida:036806} to excited states where the
singlet-triplet mixing can be neglected.

%
%
Our starting point is the Hamiltonian of two interacting electrons
in a double quantum dot with dot separation $d$ as recently applied
in studies of electron structure as well as in studies of photon
induced controlled transport.~\cite{saelen, popsueva,harju},

\begin{equation}
  \widehat{\rm{H}} = h({\mathbf{r}_1}) + h({\mathbf{r}_2}) + \frac{e^2}{4\pi\epsilon_r\epsilon_0r_{12}}
  \label{eq:hamiltonian}
\end{equation}
where
\begin{eqnarray}
  \hat h\left( x,y \right) &=& -\frac{\hbar^2}{2m^*}\nabla^2 +
  \frac{1}{2} m^*\omega^2 \left[ \left(|x| - \frac{d}{2}\right)^2 +
  y^2 \right] \nonumber \\
  &&+\frac{e^2}{8m^*}B_{ext}^2(x^2+y^2)+\frac{e}{2m^*}B_{ext}L_z\nonumber\\
  && + \gamma_eB_{ext}S_z + e\xi(t)x. \nonumber \\
   \label{eq:hamiltonian2}
\end{eqnarray}
Here $\mathbf{r}_{1,2}$ are single-particle coordinates, $\xi$ is an
electric time dependent field applied along the inter-dot axis and
$B_{ext}$ is an external magnetic field perpendicular to the dot.
The material parameters are taken as those of GaAs, with effective
mass $m^* = 0.067m_e$ and relative permittivity $\epsilon_r = 12.4$.
The gyromagnetic ratio for GaAs is $\gamma_e = g^*\frac{e}{2m_e}$,
with $g^* = -0.44$. The confinement strength is set to $\hbar\omega
= 1$~meV and the interdot separation to $d=130$~nm, which are
realistic experimental values~\cite{zumbuhl:256801,petta1}.

\begin{figure}[t]
\begin{center}
  \includegraphics[width=1\columnwidth]{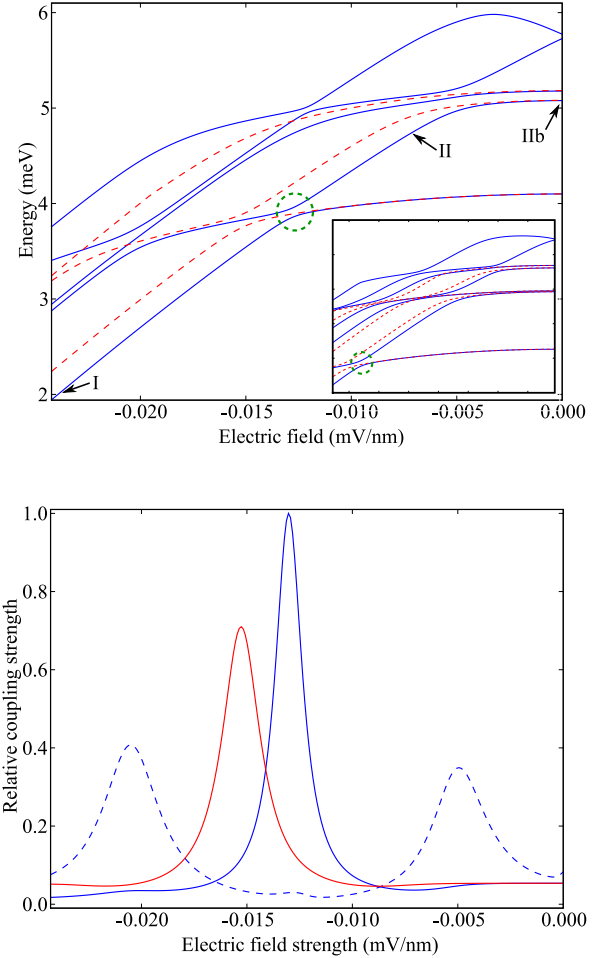}
  \caption{(color online) Upper panel shows a few of the lowest energy
  levels as a function of electric field in the $x$-direction, corresponding to $y$-symmetric states.
  Inset shows the effect of an external magnetic field (200 mT) on the spectrum.
  Lower panel shows the relative coupling strengths of selected
  states; Full blue curve is
  $\langle 1,S | X | 0,S \rangle$ between the two lowest singlet states
  (full blue curve), full red curve is $\langle 1,T_0 | X | 0,T_0 \rangle$
  between the two lowest triplet states and
  $\langle 3,S | X | 1,S \rangle $
  dashed blue curve is between the second lowest and third lowest singlet.
  Note that the $m_s = \pm 1$ triplet states are not shown.}
  \label{fig1}
\end{center}
\end{figure}

%
%
Fig.~\ref{fig1} shows the energy spectrum obtained from diagonalization with
$B_{ext}=0$ and $B_{ext}=200$~mT (inset). The spectrum of the system has been
explained in detail elsewhere~\cite{popsueva}. For $B_{ext}=0$ symmetry about
the $y$-axis is conserved and we show only the singlet and triplet ($m_s=0$) eigenstates
corresponding to the three lowest energy bands with even reflection symmetry.
This symmetry is broken when the external magnetic field is applied and all
states are shown. When $B_{ext}\neq 0$ the magnetic sub-levels split (not shown).
 In addition to the splitting of the spin states (anomalous Zeeman effect)
 the spectrum is also changed by the Zeeman ($L_z$) and diamagnetic ($B_{ext}^2$) terms,
 as seen clearly in the inset. The most important effects of these terms are the modified
singlet-triplet splitting $J$, the modified anti-crossing energy
difference and the splitting of the  second band according to the
sign of the angular momentum expectation value, $\left<L_z\right>$.

It is particularly worth noting that the energy spectrum exhibit
several near degenerate anti-crossing regions where the coupling
strength can be tuned in experiments through adjustable gate
voltages and switching times~\cite{laird}. Restricted by
conservation of total spin, the states can couple dynamically as the
electric field varies. The relative coupling strength from our model
is shown in the lower panel of Fig.~\ref{fig1}. The strongest
coupling strength is seen between the singlet ground state and first
excited states at $-0.013$~mV/nm. This puts a limit on the switching
time through the region enclosed by a green circle for adiabatic
time development along the initial singlet ground state,  marked as
'I' in Fig.~\ref{fig1}. On the other hand, a very rapid transfer can
lead to diabatic development, to be discussed later.

%
%
The few and well defined states resulting from the present
diagonalization suggest that the time evolution is most precisely
described in an adiabatic basis of instant eigenstates of the
time-dependent Hamiltonian of Eq.~{(\ref{eq:hamiltonian})},
$\widehat{H}(t)\chi(\mathbf{r}_1,\mathbf{r}_2;\xi) =
\epsilon(\xi)\chi(\mathbf{r}_1,\mathbf{r}_2;\xi)$, where the
energies $\epsilon(\xi)$ depend parametrically on the time-dependent
electric field. We expand the wavefunction in these basis
states,
\begin{equation}
\Psi (\mathbf{r}_1,\mathbf{r}_2,t) = \sum_{k}
c_{k}(t)\chi_k(\mathbf{r}_1,\mathbf{r}_2;\xi) \otimes \ket{S}
\label{eq:psi}
\end{equation}
where $k$ runs over all basis states. Projecting onto each basis
state, the following expression for the time evolution of the
amplitudes is obtained,
\begin{equation}
\dot c_k(t)\! = \dot{\xi} \ \sum_{j \neq k} \frac{\bra{\chi_k} X
\ket{\chi_j}} {\epsilon_k-\epsilon_j}c_j(t) +
\imath\epsilon_k(\xi)c_k(t),
\end{equation}
with $X = x_1 + x_2$.

%
%
Additional terms can readily be included as extra matrix elements in
the expression above. To study hyperfine interactions, spin
couplings for $\sim 10^6$ nuclear spins surrounding the electrons
must be included. These evolve in time, but on a much longer time
scale than we will consider here. We therefore use the quasistatic
approximation~\cite{PhysRevB.65.205309, taylor}, neglecting their
time dependence. In addition, the large number of spins justifies a
semiclassical approximation~\cite{briggs-rost}, where all the
nuclear spins are described by a single classical magnetic field.
The hyperfine interaction is then given by
\begin{equation}
 \widehat{H}_{N} = \gamma_e \sum_{i=1,2} {\bf S}_i \cdot {\bf B}_N (r_i) \label{HSNC},
\end{equation}
where $S_i$ is the spin operator of electron $i$. Generally the
direction of nuclear magnetic field $\bm{B}_N$ is random (no
polarization) and the magnitude varies according to a normal
distribution about zero, $ P(\bm{B}_N) = 1/(2\pi
B^2_{\rm{nuc}})^{\frac{3}{2}}
\exp\left(-\bm{B}_N\cdot\bm{B}_N/2B^2_{\rm{nuc}}\right)$~\cite{loss}.
$B_{\rm nuc}$ can be determined by experiments and is of the order
of $1$~mT~\cite{petta1}. The precise spatial variation of the
nuclear magnetic field $\bm{B}_N$ is in general unknown, but also of
less importance. The essential feature in the spin dephasing
mechanism is the difference in effective magnetic fields between the
two dots. The simplest way to represent this is by a step function,
\begin{equation}
    {\bf B}_N =
    \begin{cases}
       \left( B_x \hat{{\bf e}}_{x} \ + \ B_y \hat{{\bf e}}_{y} + \ B_z \hat{{\bf e}}_{z}\right) & \text{, for } x \ge 0 \\
        0 & \text{, otherwise }
    \end{cases}
    \label{eq:Bn}
\end{equation}
This induces coupling between the singlet and triplet
states and between the different triplet states.

%
%
From Eqs.~(\ref{eq:hamiltonian},\ref{eq:psi},\ref{eq:Bn}) the time
evolution of the wavefunction at zero electric field, restricted to
the four lowest energy states $\{\ket{S}, \ket{T_0},
\ket{T_-},\ket{T_+}\}$, becomes,
\begin{equation}
    \mathbf{\dot c}(t) = \imath\gamma_e\left(
    \begin{matrix}
        J/\gamma_e & B_z & \frac{B_x - \imath B_y}{\sqrt{2}} & -\frac{B_x + \imath B_y}{\sqrt{2}} \\
        B_z & 0 & 0 & 0 \\
        \frac{B_x + \imath B_y}{\sqrt2} & 0 & -B_{ext} & 0 \\
        -\frac{B_x - \imath B_y}{\sqrt2} & 0 & 0 & B_{ext}
    \end{matrix}
    \right)\mathbf{c}(t).
\label{eq:matrise}
\end{equation}
This expression is identical to four-level models previously
considered~\cite{taylor}. We have here, however, excluded the
inter-triplet couplings as this will allow us to obtain an
analytical solution. When the external magnetic field is
sufficiently strong, the $m_s=\pm 1$ triplet components effectively
decouple, and we are left with a two-level system defined by the
upper left part of the four-level matrix. Furthermore, since the
triplet sub-levels are degenerate, the four-level matrix for $B_{\rm
ext}=0$ may be represented by an effective "radial" two-level model.
In both cases we obtain the time evolution of the singlet
coefficient
\begin{equation}
    |c_S(t)|^2 = 1-\frac{4B^2}{4B^2+J^2}\sin^2{\left(\frac{1}{2}t\sqrt{4B^2+J^2}\right)},
\label{c_S}
\end{equation}
with $B = B_z$ for $B_{ext} \gg J$ and $B = \sqrt{B_x^2+B_y^2+B_z^2}$ for $B_{ext}=0$. \\

%
%
\begin{figure}[t]
\begin{center}
    \includegraphics[width=1\columnwidth]{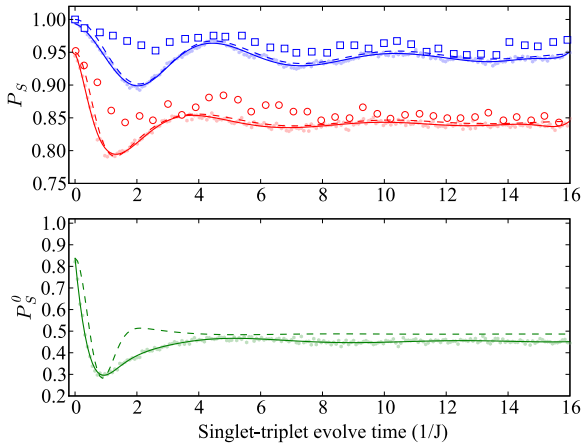}
    \caption{(color online) Upper panel: Averaged time evolution of singlet component with $B_{ext} = 200$~mT.
             $B_{N}/J = 0.4$ (full blue line) and $B_{N}/J = 0.91$ (full red line, downshifted 5\%).
             Squares and circles are experimental data taken from~\cite{laird}.
             Prediction from two-level model shown as dashed lines.
             Semitransparent dots are actual numerical data, full lines
             are obtained from a smoothed spline interpolation.
             Lower panel:  $B_{N}/J = 0.91$, but without external magnetic field.
             Prediction from four-level model (dashed line), scaled to match at $t=0$.
    }
  \label{fig2}
\end{center}
\end{figure}

In Fig.~\ref{fig2} we show our results together with
experimental results of Laird et. al~\cite{laird} using $B_{ext}=200$~mT (upper panel).
To calculate the singlet-triplet dephasing we start out in the singlet
ground state at large electric field, $-0.024$~mV/nm.
The field is then switched adiabatically to zero within 1 ns and kept at zero a
variable period of time, $t_s$ up to $16 \ J/\hbar$ ns, before being switched
back to its original value. The procedure is repeated a number of times, with
a random nuclear magnetic field drawn from a Gaussian distribution. Finally,
the average of the singlet correlator is computed, $P_S = 1/N\sum_i|c^i_S(t_s)|^2$.
The full lines in the upper ($B_{\rm ext} = 200$~mT) and lower panel ($B_{\rm ext}=0$)
of Fig.~\ref{fig2} are produced from a sample size of $N = 1000$ different nuclear fields. For clarity, we have
interpolated the numerical data using splines with a smoothing requirement.
The actual numerical points are shown as semi-transparent dots. Also
shown, as dashed lines, are predictions from the two-level theory.
An excellent agreement between the theoretical results are noted
which indicate that effects of the electrical switch, excited states
and geometry of the potential are of less importance in this
case. The present results are also compared with experimental data,
shown as dots and circles in the upper panel. We also observe a
very good agreement with the experimental results.
Here we have varied $B_{\rm nuc}$ as opposed to $J$ which may be varied
in experiments by tuning the gate voltages. As verified by the two-level models
the dephasing process mainly relies on their ratio. It should be noted that results in the upper panel have
been scaled according to $P_S(t) = 1-V(1-P^0_S(t))$, where $V =
0.40$ is a visibility parameter determined in the experiments,
$P_S(t)$ is the experimental averaged correlator while $P^0_S(t)$ is
the theoretical averaged correlator~\cite{laird}.

Setting the external magnetic field to zero leads to increased
dephasing, since the $m_s = \pm 1$ states are now coupled to the
initial singlet state. This is indeed what we observe in our
simulations, as shown in the lower panel of Fig.~\ref{fig2}. The
four-level (effective two-level) model yields the result shown as a
dashed line, scaled to match the numerical data at $t_s=0$. At large
evolve times, these are in good agreement. Numerical solution of the
full four-level matrix suggests that the slight discrepancy around
$t_s = 2$ is due to the neglected inter-triplet couplings in
Eq.~{(\ref{eq:matrise})}. In contrast to the two-level case, we
observe in the four level case a $\sim$10\% dephasing occurring
during the switch or more precisely between the two passages of the
avoided crossing circled out in Fig.~\ref{fig1}. This is a result of
a much more involved dynamical interplay between the states and
result in the present theory to a reduction of $P_S^0$ at $t_s = 0$.

\begin{figure}[t]
\begin{center}
  \includegraphics[width=1\columnwidth]{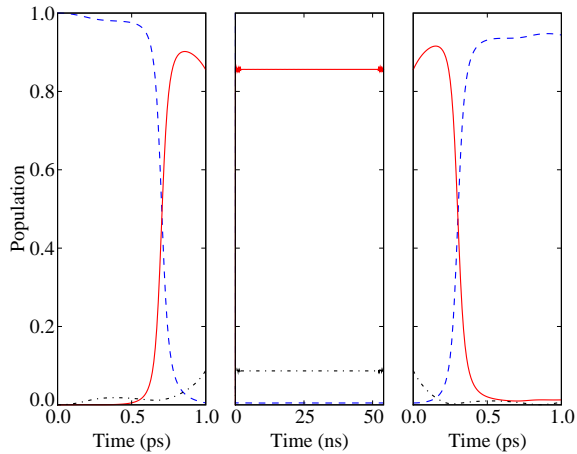}
  \caption{(color online) The time evolution of the three most prominent states during the fast switch proceedure. About $95$ \% of the total norm is represented by these states, which are \ket{S,0} (blue dashed), \ket{S,1} (red full) and \ket{S,2} (black dash-dotted). $B_{nuc} = 1$~mT ([1,1,1]). $B_{ext} = 0$. Left panel shows evolution during rapid switch ($1$~ps), center panel shows adiabatic passage to first excited state and evolution during dephasing period. Right panel shows rapid switching back to the one-well configuration ($1$~ps). Singlet-triplet coupling is weak for the excited singlet state, and thus only $\sim 5$ \% is lost (mainly to other singlet states during switching).}
  \label{fig3}
\end{center}
\end{figure}

%
%
A completely new feature can be studied by introducing an ultrafast
switching function which leads to diabatic passage through the
anti-crossing, highlighted by the green circle in Fig.~\ref{fig1}.
This will transfer the system to the first excited singlet state
where at zero electric field the singlet-triplet splitting is
approximately 100 times greater compared to the ground state.
Important to note is that states in the second energy band are like
the ground state covalent states~\cite{popsueva} with one electron
in each dot. This makes the states well suited for single-electron
gate operations unlike the states in the third energy band which
resemble ionic states. In Fig.~\ref{fig3} the evolution of the three
lowest energy singlet states are displayed as the system is rapidly
switched from positions marked I to II in Fig.~\ref{fig1}, using a
nuclear magnetic field of $1$~mT, and zero external magnetic field.
From there it is transferred adiabatically from II to IIb. The
leftmost panel shows the transfer of population from the ground to
first excited state during a $1$~ps switch, with around 10\% of the
population being further transported to the second excited state. In
the middle panel, the system is switched adiabatically to zero
electric field during $2$~ns, and left to evolve here for $50$~ns.
The singlet-triplet coupling is completely suppressed, and when the
system is switched back to position I, 95\% of the initial singlet
population is regained, the rest having vanished mostly to higher
excited states during passage through the anti-crossings. The
simulation was repeated a number of times with increasing nuclear
magnetic field strength up to $10$~mT, attributing only negligible
changes to the dynamics. We note that by applying optimal control
schemes the transition to excited states may be achieved with near
100\% transition probability~\cite{murgida:036806, saelen,
rasanen:157404}.

%
%
In conclusion, we have applied a first principle molecular orbital
framework to accurately calculate the time development of the
electronic states of a two-electron quantum dot molecule. For the
first time the hyperfine interaction has been taken into account on
the same coherent level as the time variation of the external
electromagnetic fields. Calculations have displayed a very good
agreement with experiments and a previously developed two-level
model for experiments performed in static external magnetic fields.
We have also performed full numerical calculations in absence of
external magnetic field and predicted the dephasing dynamics by an
effective four level model. Finally, we have pointed towards a new
solution to the spin-dephasing problem by controlling and applying
transitions between excited states.

This research has been supported by the Research Council of Norway
(RCN). LS acknowledge financially support through a mobility grant
from Universit\' e Pierre et Marie Curie.

%
%


\begin{thebibliography}{18}
\expandafter\ifx\csname natexlab\endcsname\relax\def\natexlab#1{#1}\fi
\expandafter\ifx\csname bibnamefont\endcsname\relax
  \def\bibnamefont#1{#1}\fi
\expandafter\ifx\csname bibfnamefont\endcsname\relax
  \def\bibfnamefont#1{#1}\fi
\expandafter\ifx\csname citenamefont\endcsname\relax
  \def\citenamefont#1{#1}\fi
\expandafter\ifx\csname url\endcsname\relax
  \def\url#1{\texttt{#1}}\fi
\expandafter\ifx\csname urlprefix\endcsname\relax\def\urlprefix{URL }\fi
\providecommand{\bibinfo}[2]{#2}
\providecommand{\eprint}[2][]{\url{#2}}

\bibitem[{\citenamefont{Overhauser}(1953)}]{overhauser}
\bibinfo{author}{\bibfnamefont{A.~W.} \bibnamefont{Overhauser}},
  \bibinfo{journal}{Phys. Rev.} \textbf{\bibinfo{volume}{92}},
  \bibinfo{pages}{411} (\bibinfo{year}{1953}).

\bibitem[{\citenamefont{Khaetskii et~al.}(2002)\citenamefont{Khaetskii, Loss,
  and Glazman}}]{khaetskii}
\bibinfo{author}{\bibfnamefont{A.~V.} \bibnamefont{Khaetskii}},
  \bibinfo{author}{\bibfnamefont{D.}~\bibnamefont{Loss}}, \bibnamefont{and}
  \bibinfo{author}{\bibfnamefont{L.}~\bibnamefont{Glazman}},
  \bibinfo{journal}{Phys. Rev. Lett.} \textbf{\bibinfo{volume}{88}},
  \bibinfo{pages}{186802} (\bibinfo{year}{2002}).

\bibitem[{\citenamefont{Koppens~\textit{et al.}}(2006)}]{koppens1}
\bibinfo{author}{\bibfnamefont{F.~H.~L.} \bibnamefont{Koppens~\textit{et
  al.}}}, \bibinfo{journal}{Nature} \textbf{\bibinfo{volume}{442}},
  \bibinfo{pages}{766} (\bibinfo{year}{2006}).

\bibitem[{\citenamefont{Petta et~al.}(2004)\citenamefont{Petta, Johnson,
  Marcus, Hanson, and Gossard}}]{petta1}
\bibinfo{author}{\bibfnamefont{J.~R.} \bibnamefont{Petta}},
  \bibinfo{author}{\bibfnamefont{A.~C.} \bibnamefont{Johnson}},
  \bibinfo{author}{\bibfnamefont{C.~M.} \bibnamefont{Marcus}},
  \bibinfo{author}{\bibfnamefont{M.~P.} \bibnamefont{Hanson}},
  \bibnamefont{and} \bibinfo{author}{\bibfnamefont{A.~C.}
  \bibnamefont{Gossard}}, \bibinfo{journal}{Phys. Rev. Lett.}
  \textbf{\bibinfo{volume}{93}}, \bibinfo{pages}{186802}
  (\bibinfo{year}{2004}).

\bibitem[{\citenamefont{Loss and DiVincenzo}(1998)}]{divincenzo}
\bibinfo{author}{\bibfnamefont{D.}~\bibnamefont{Loss}} \bibnamefont{and}
  \bibinfo{author}{\bibfnamefont{D.~P.} \bibnamefont{DiVincenzo}},
  \bibinfo{journal}{Phys. Rev. A} \textbf{\bibinfo{volume}{57}},
  \bibinfo{pages}{120} (\bibinfo{year}{1998}).

\bibitem[{\citenamefont{Koppens~\textit{et al.}}(2005)}]{koppens2}
\bibinfo{author}{\bibfnamefont{F.~H.~L.} \bibnamefont{Koppens~\textit{et
  al.}}}, \bibinfo{journal}{Science} \textbf{\bibinfo{volume}{309}},
  \bibinfo{pages}{1346} (\bibinfo{year}{2005}).

\bibitem[{\citenamefont{Petta~\textit{et al.}}(2005)}]{petta2}
\bibinfo{author}{\bibfnamefont{J.~R.} \bibnamefont{Petta~\textit{et al.}}},
  \bibinfo{journal}{Science} \textbf{\bibinfo{volume}{309}},
  \bibinfo{pages}{2180} (\bibinfo{year}{2005}).

\bibitem[{\citenamefont{Taylor et~al.}(2007)\citenamefont{Taylor, Petta,
  Johnson, Yacoby, Marcus, and Lukin}}]{taylor}
\bibinfo{author}{\bibfnamefont{J.~M.} \bibnamefont{Taylor}},
  \bibinfo{author}{\bibfnamefont{J.~R.} \bibnamefont{Petta}},
  \bibinfo{author}{\bibfnamefont{A.~C.} \bibnamefont{Johnson}},
  \bibinfo{author}{\bibfnamefont{A.}~\bibnamefont{Yacoby}},
  \bibinfo{author}{\bibfnamefont{C.~M.} \bibnamefont{Marcus}},
  \bibnamefont{and} \bibinfo{author}{\bibfnamefont{M.~D.} \bibnamefont{Lukin}},
  \bibinfo{journal}{Phys. Rev. B} \textbf{\bibinfo{volume}{76}},
  \bibinfo{pages}{035315} (\bibinfo{year}{2007}).

\bibitem[{\citenamefont{Coish and Loss}(2005)}]{loss}
\bibinfo{author}{\bibfnamefont{W.~A.} \bibnamefont{Coish}} \bibnamefont{and}
  \bibinfo{author}{\bibfnamefont{D.}~\bibnamefont{Loss}},
  \bibinfo{journal}{Phys. Rev. B} \textbf{\bibinfo{volume}{72}},
  \bibinfo{pages}{125337} (\bibinfo{year}{2005}).

\bibitem[{\citenamefont{Laird et~al.}(2006)\citenamefont{Laird, Petta, Johnson,
  Marcus, Yacoby, Hanson, and Gossard}}]{laird}
\bibinfo{author}{\bibfnamefont{E.~A.} \bibnamefont{Laird}},
  \bibinfo{author}{\bibfnamefont{J.~R.} \bibnamefont{Petta}},
  \bibinfo{author}{\bibfnamefont{A.~C.} \bibnamefont{Johnson}},
  \bibinfo{author}{\bibfnamefont{C.~M.} \bibnamefont{Marcus}},
  \bibinfo{author}{\bibfnamefont{A.}~\bibnamefont{Yacoby}},
  \bibinfo{author}{\bibfnamefont{M.~P.} \bibnamefont{Hanson}},
  \bibnamefont{and} \bibinfo{author}{\bibfnamefont{A.~C.}
  \bibnamefont{Gossard}}, \bibinfo{journal}{Phys. Rev. Lett.}
  \textbf{\bibinfo{volume}{97}}, \bibinfo{pages}{056801}
  (\bibinfo{year}{2006}).

\bibitem[{\citenamefont{Murgida et~al.}(2007)\citenamefont{Murgida, Wisniacki,
  and Tamborenea}}]{murgida:036806}
\bibinfo{author}{\bibfnamefont{G.~E.} \bibnamefont{Murgida}},
  \bibinfo{author}{\bibfnamefont{D.~A.} \bibnamefont{Wisniacki}},
  \bibnamefont{and} \bibinfo{author}{\bibfnamefont{P.~I.}
  \bibnamefont{Tamborenea}}, \bibinfo{journal}{Phys. Rev. Lett.}
  \textbf{\bibinfo{volume}{99}}, \bibinfo{eid}{036806} (\bibinfo{year}{2007}).

\bibitem[{\citenamefont{Saelen et~al.}()\citenamefont{Saelen, Nepstad, Degano,
  and Hansen}}]{saelen}
\bibinfo{author}{\bibfnamefont{L.}~\bibnamefont{Saelen}},
  \bibinfo{author}{\bibfnamefont{R.}~\bibnamefont{Nepstad}},
  \bibinfo{author}{\bibfnamefont{I.}~\bibnamefont{Degano}}, \bibnamefont{and}
  \bibinfo{author}{\bibfnamefont{J.~P.} \bibnamefont{Hansen}},
  \bibinfo{howpublished}{Subm. to Phys. Rev. Lett. 2007}.

\bibitem[{\citenamefont{Popsueva et~al.}(2007)\citenamefont{Popsueva, Nepstad,
  Birkeland, F{\o}rre, Hansen, Lindroth, and Walterson}}]{popsueva}
\bibinfo{author}{\bibfnamefont{V.}~\bibnamefont{Popsueva}},
  \bibinfo{author}{\bibfnamefont{R.}~\bibnamefont{Nepstad}},
  \bibinfo{author}{\bibfnamefont{T.}~\bibnamefont{Birkeland}},
  \bibinfo{author}{\bibfnamefont{M.}~\bibnamefont{F{\o}rre}},
  \bibinfo{author}{\bibfnamefont{J.~P.} \bibnamefont{Hansen}},
  \bibinfo{author}{\bibfnamefont{E.}~\bibnamefont{Lindroth}}, \bibnamefont{and}
  \bibinfo{author}{\bibfnamefont{E.}~\bibnamefont{Walterson}},
  \bibinfo{journal}{Phys. Rev. B} \textbf{\bibinfo{volume}{76}},
  \bibinfo{pages}{035303} (\bibinfo{year}{2007}).

\bibitem[{\citenamefont{Harju et~al.}(2002)\citenamefont{Harju, Siljamaki, and
  Nieminen}}]{harju}
\bibinfo{author}{\bibfnamefont{A.}~\bibnamefont{Harju}},
  \bibinfo{author}{\bibfnamefont{S.}~\bibnamefont{Siljamaki}},
  \bibnamefont{and} \bibinfo{author}{\bibfnamefont{R.~M.}
  \bibnamefont{Nieminen}}, \bibinfo{journal}{Phys. Rev. Lett.}
  \textbf{\bibinfo{volume}{88}}, \bibinfo{pages}{226804}
  (\bibinfo{year}{2002}).

\bibitem[{\citenamefont{Zumb\"{u}hl et~al.}(2004)\citenamefont{Zumb\"{u}hl,
  Marcus, Hanson, and Gossard}}]{zumbuhl:256801}
\bibinfo{author}{\bibfnamefont{D.~M.} \bibnamefont{Zumb\"{u}hl}},
  \bibinfo{author}{\bibfnamefont{C.~M.} \bibnamefont{Marcus}},
  \bibinfo{author}{\bibfnamefont{M.~P.} \bibnamefont{Hanson}},
  \bibnamefont{and} \bibinfo{author}{\bibfnamefont{A.~C.}
  \bibnamefont{Gossard}}, \bibinfo{journal}{Phys. Rev. Lett.}
  \textbf{\bibinfo{volume}{93}}, \bibinfo{pages}{256801}
  (\bibinfo{year}{2004}).

\bibitem[{\citenamefont{Merkulov et~al.}(2002)\citenamefont{Merkulov, Efros,
  and Rosen}}]{PhysRevB.65.205309}
\bibinfo{author}{\bibfnamefont{I.~A.} \bibnamefont{Merkulov}},
  \bibinfo{author}{\bibfnamefont{A.~L.} \bibnamefont{Efros}}, \bibnamefont{and}
  \bibinfo{author}{\bibfnamefont{M.}~\bibnamefont{Rosen}},
  \bibinfo{journal}{Phys. Rev. B} \textbf{\bibinfo{volume}{65}},
  \bibinfo{pages}{205309} (\bibinfo{year}{2002}).

\bibitem[{\citenamefont{Briggs and Rost}(2000)}]{briggs-rost}
\bibinfo{author}{\bibfnamefont{J.~S.} \bibnamefont{Briggs}} \bibnamefont{and}
  \bibinfo{author}{\bibfnamefont{J.~M.} \bibnamefont{Rost}},
  \bibinfo{journal}{Eur. Phys. Journ. D} \textbf{\bibinfo{volume}{10}},
  \bibinfo{pages}{311} (\bibinfo{year}{2000}).

\bibitem[{\citenamefont{R\"{a}s\"{a}nen
  et~al.}(2007)\citenamefont{R\"{a}s\"{a}nen, Castro, Werschnik, Rubio, and
  Gross}}]{rasanen:157404}
\bibinfo{author}{\bibfnamefont{E.}~\bibnamefont{R\"{a}s\"{a}nen}},
  \bibinfo{author}{\bibfnamefont{A.}~\bibnamefont{Castro}},
  \bibinfo{author}{\bibfnamefont{J.}~\bibnamefont{Werschnik}},
  \bibinfo{author}{\bibfnamefont{A.}~\bibnamefont{Rubio}}, \bibnamefont{and}
  \bibinfo{author}{\bibfnamefont{E.~K.~U.} \bibnamefont{Gross}},
  \bibinfo{journal}{Phys. Rev. Lett.} \textbf{\bibinfo{volume}{98}},
  \bibinfo{eid}{157404} (\bibinfo{year}{2007}).

\end{thebibliography}
\end{document}